\documentclass[10pt]{article}
\usepackage[a4paper]{geometry}
\usepackage[utf8]{inputenc}
\usepackage[T1]{fontenc}
\usepackage{setspace,authblk}
\usepackage{amsmath,mathrsfs,amsfonts,amsthm,amssymb}

\usepackage{tikz}

\usepackage[english]{babel}

\usepackage{graphicx}
 
\usepackage{hyperref}

\title{\singlespace {Knowing that you do not know everything}}

\author[]{Alex A.T. Rathke\thanks{NECCT/FEA-RP/USP, University of S\~ao Paulo. \texttt{alex.rathke@alumni.usp.br}}}

\date{\today}

\theoremstyle{plain}
\newtheorem{theorem}{Theorem}
\newtheorem*{theorem*}{Theorem}

\newtheorem*{proposition*}{Proposition}
\newtheorem{remark}{Remark}
\newtheorem*{remark*}{Remark}

\newtheorem*{condition*}{Condition}

\newtheorem*{definition*}{Definition}

\newtheorem*{assumption*}{Assumption}

\begin{document}

\maketitle

\begin{abstract}

We show that a rational agent with true and refinable knowledge of events cannot know if she knows everything or not. This epistemic limitation is not resolved by introspection about tautologies or by learning about new events.

\end{abstract}


\section{Introduction} \label{Introduction}

Epistemic models in game theory and economics incorporate formal representations of the knowledge (and belief) of agents into decision models. In the usual approach, an algebra of events represents the information structure of the states of the world. A rational agent applies an epistemic mechanism to access the information included in each event, which provides a subset of states representing the information that the agent is capable to process, given the event of interest.

Studies propose several axiomatic properties of the knowledge of an agent. The essential property of knowledge derived from modal logic is that knowledge is closed under logical consequence (\cite{kripke1959,fagin1987,samet1990,fagin2003}). It means that if the agent knows an implication, then knowing the premise implies that she also knows the conclusion. In the set-theoretic framework, this property operates as a monotonic mapping which preserves set-inclusion order, therefore preserving refinements. We refer to the property of \emph{Monotonicity}.

Knowledge also satisfies the condition of always being true (\cite{kripke1959,hintikka1962,kripke1963,fagin1987,samet1990,fagin2003}), as opposite to belief, which may be true or false. It means that an agent knows that an event obtains if and only if her knowledge is a subset which includes only states within that same event\footnote{In modal logic and epistemic logic, closure under logical consequence is usually referred as the \emph{axiom K}, and knowledge always being true is referred as the \emph{axiom T} (\cite{kripke1959,hintikka1962,kripke1963,fagin2003}).}. We refer to the \emph{axiom of Truth}.

Assuming Truth and Monotonicity, we show that a rational agent cannot know if she knows every state of the world or not. Introspection about everything that she either knows or that she does not know provides just the same knowledge that she already has. Learning about new events provides information about the lack of knowledge prior to learning them, however this new information is not sufficient for the agent to know that any other unknown events either exist or not. We present our analysis and discussion in the following Sections.

\section{Analysis} \label{Analysis}

Assume a set of states $\Omega$ and a complete algebra of events $\mathcal{E} \subseteq 2^{\Omega}$ which includes $\Omega$ and is closed under complementation and arbitrary unions, hence closed under arbitrary intersections as well. Define an epistemic operator equal to $K : \mathcal{E} \rightarrow \mathcal{E}$ which is a mapping across events $E \in \mathcal{E}$. The set $K(E) = KE$ represents the event that the agent knows that the event $E$ obtains. The complement set $\neg KE = \Omega \setminus KE$ is the event that the agent does not know that $E$ obtains. Iterations of the epistemic operator $K$ model the introspection process of the agent, e.g. the set $K(\neg KE) = K \neg KE$ is the event that the agent knows that she does not know that $E$ obtains\footnote{We apply the following notational convention, which is usual in the literature: for any subsets $E,F \in \mathcal{E}$, we regard the complement $\neg E = \Omega \setminus E$ followed by the relative complement $E \setminus F = E \cap \neg F$, in this order, then we regard the other set operations. For example, we apply the unambiguous notation $KE \cup \neg F \setminus E = (KE) \cup ((\neg F) \setminus E )$.}.

For all events $E,F \in \mathcal{E}$, we assume that the operator $K$ represents the knowledge of a rational agent if it satisfies at least two main conditions. First, the \emph{axiom of Truth} equal to $KE \subseteq E$ states that the agent's knowledge of $E$ must include only states within that event. Otherwise, she would regard states within the negation $\neg E$ as true, although these states are false when $E$ obtains\footnote{An operator on $E$ for which the image includes states within the complement $\neg E$ is commonly applied to model the belief of agents (\cite{hintikka1962,fagin1987,fagin2003,heifetz2006,schipper2014,fukuda2019}). A belief operator do not necessarily satisfy the axiom of Truth, since agents may have false beliefs.}. Second, the property of \emph{Monotonicity} states that $E \subseteq F$ implies $KE \subseteq KF$. It indicates that refined events imply refined knowledge, and it provides a way for the agent to make logical inferences (\cite{kripke1959,hintikka1962,kripke1963,fagin1987,samet1990,fagin2003,heifetz2006,li2009,samet2010,galanis2013,schipper2014,fukuda2019,tada2024})\footnote{Studies on epistemic game theory and decision theory usually combine Truth and Monotonicity with additional assumptions of knowledge. The main properties assumed in the literature are the following, for all events $E \in \mathcal{E}$ (\cite{kripke1959,hintikka1962,kripke1963,fagin1987,samet1990,fagin2003,heifetz2006,li2009,samet2010,galanis2013,schipper2014,fukuda2019,tada2024}):

\noindent\emph{Axiom of Necessitation:} $K \Omega = \Omega$;

\noindent\emph{Positive Introspection:} $KE \subseteq KKE$;

\noindent\emph{Negative Introspection:} $\neg KE \subseteq K \neg KE$. }.

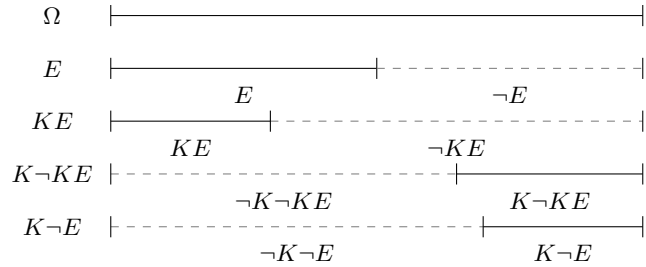
\begin{figure}[htb!]
\begin{center}
\begin{tikzpicture}[scale=0.7,font=\small] 
	\draw (0,10) -- (10,10); 
	\draw (0,10.2) -- (0,9.8); 
	\draw (10,10.2) -- (10,9.8); 
	\node (omega) at (-1.1,10) {$\Omega$}; 
	\draw (0,9) -- (5,9); 
	\draw (0,9.2) -- (0,8.8); 
	\draw (10,9.2) -- (10,8.8); 
	\draw (5,9.2) -- (5,8.8); 
	\draw[gray, dashed] (5,9) -- (10,9); 
	\node (levent) at (-1.1,9) {$E$}; 
	\node (event) at (2.5,8.5) {$E$}; 
	\node (negevent) at (7.5,8.5) {$\neg E$}; 
	\draw (0,8) -- (3,8); 
	\draw[gray, dashed] (3,8) -- (10,8); 
	\node (lkevent) at (-1.1,8) {$KE$}; 
	\draw (0,8.2) -- (0,7.8); 
	\draw (10,8.2) -- (10,7.8); 
	\draw (3,8.2) -- (3,7.8); 
	\node (kevent) at (1.5,7.5) {$KE$}; 
	\node (negkevent) at (6.5,7.5) {$\neg KE$}; 
	\draw[gray, dashed] (0,7) -- (6.5,7); 
	\draw (6.5,7) -- (10,7); 
	\node (lknegkevent) at (-1.1,7) {$K \neg KE$}; 
	\draw (0,7.2) -- (0,6.8); 
	\draw (10,7.2) -- (10,6.8); 
	\draw (6.5,7.2) -- (6.5,6.8); 
	\node (knegkevent) at (8.25,6.5) {$K \neg KE$}; 
	\node (negknegkevent) at (3.25,6.5) {$\neg K \neg KE$}; 
	\draw[gray, dashed] (0,6) -- (7,6); 
	\draw (7,6) -- (10,6); 
	\node (lknegevent) at (-1.1,6) {$K \neg E$}; 
	\draw (0,6.2) -- (0,5.8); 
	\draw (10,6.2) -- (10,5.8); 
	\draw (7,6.2) -- (7,5.8); 
	\node (knegevent) at (8.5,5.5) {$K \neg E$}; 
	\node (negknegevent) at (3.5,5.5) {$\neg K \neg E$}; 
\end{tikzpicture}
\end{center}
\caption{An agent's knowledge of an event, $KE$.} \label{ss01}
\end{figure}

Regard Fig. \ref{ss01} where we represent a general case with a state-space $\Omega$, an event $E$ and the agent's knowledge of the event, $KE$. The event of not knowing $E$ is equal to $\neg KE$. The agent may think about her lack of knowledge of $E$, which provides the knowledge that she does not know that $E$ obtains, equal to $K \neg KE$. Notice that this event $K \neg KE$ is different than the event of knowing that $E$ does not obtain, which is equal to $K \neg E$. Since the axiom of Truth implies $\neg E \subseteq \neg KE$, then Monotonicity implies $K \neg E \subseteq K \neg KE$, see Fig. \ref{ss01}. For the agent, it indicates that knowing that $E$ does not obtain ($K\neg E$) is at least as a refined knowledge as knowing that she does not know that $E$ obtains ($K \neg KE$). For $E \neq \Omega$, we find the general conditions equal to

\begin{equation} \label{refine01}
\begin{array}{c}
K \neg E \subseteq \neg E \subseteq \neg KE \not\subset E, \\
\\
K \neg E \subseteq K \neg KE \subseteq \neg KE \not\subset E. \\
\end{array}
\end{equation}


All that the agent knows is bounded by her epistemic operator $K$. It means that the event $KE$ is the agent's knowledge of the event $E$, however the event $\neg KE$ does not represent the agent's knowledge, since it is not the image of $K$ over the event of interest. In special, the event $\neg KE = \neg E \cup E \setminus KE$ represents the agent's lack of knowledge of $E$, for it refers either that the event $E$ does not obtain ($\neg E$), or that the agent has limited capacity to process the full information that $E$ obtains ($E \setminus KE$), see Fig. \ref{ss01}. Complete lack of knowledge of some event $F \in \mathcal{E}$ is represented as $KF = \emptyset$, which means that there is not any state in $\Omega$ in which the agent knows that $F$ obtains.

Invoking logical omniscience, the agent may further think that any event either obtains or it does not obtain, $E \cup \neg E$, or she may introspect about that she either knows $E$ or she does not know it, $KE \cup \neg KE$, all referring to the same tautological event $E \cup \neg E = KE \cup \neg KE = \Omega$. The knowledge provided by thinking about the full state-space is equal to $K \Omega$. Notice that Truth and Monotonicity alone do not imply that her knowledge $K \Omega$ is necessarily equal to $\Omega$. We rather find the general condition $K \Omega \subseteq \Omega$, for which the non-empty complement $\neg K \Omega \neq \emptyset$ may hold. The event $\neg K \Omega$ represents the agent's limited capacity to process the full information contained in the state-space $\Omega$.

It is helpful to remind here that the equality $K \Omega = \Omega$ is called \emph{Necessitation}, for it is a condition often assumed in models of knowledge in the literature, see e.g. \cite{kripke1959,hintikka1962,kripke1963,fagin1987,samet1990,fagin2003,heifetz2006,li2009,samet2010,galanis2013,schipper2014,fukuda2019,tada2024}. We do not assume Necessitation as a primitive in our analysis\footnote{Two comments apply. First, in set-theoretic models, Necessitation implies that the agent knows the full information about every state in $\Omega$, for it is an ad hoc primitive because it is independent from other axioms of modal logic and epistemic logic. If we assume Necessitation in our analysis, we would assume \emph{ex ante} that the agent has no lack of knowledge of any state in $\Omega$ prior to any thinking or introspection. Second, there must be no confusion between logical omniscience, which derives from the law of the excluded middle in classical logic and it means that either $E$ obtains or its negation obtains, vs. Necessitation, which derives from the original interpretation of the square operator in modal logic and it means that if a proposition is true, then it is necessarily true.}, i.e. we may admit a non-empty complement $\neg K \Omega \neq \emptyset$, meaning that the agent does not know some event. Truth and Monotonicity imply the condition $\neg K \Omega \subseteq \neg K E$ for all events $E \in \mathcal{E}$

\begin{remark} \label{t01}
Regard the knowledge of an event, $KE$, its complement, $\neg KE$, and the relative complement with respect to the event $E$, equal to $E \setminus KE$. For the knowledge of the full state-space, $K \Omega$, the complement and the relative complement are the same by definition, $\neg K \Omega = \Omega \setminus K \Omega$.
\begin{proof}
The complement of $KE$ is equal to $\neg KE = \neg E \cup E \setminus KE$. Let $E = \Omega$, which provides $\neg K \Omega = \Omega \setminus K \Omega$.
\end{proof}
\end{remark}

Eq. \ref{refine01} provides the general condition $\neg KE \not\subset E$ for all events $E \neq \Omega$. Remark \ref{t01} derives a special condition equal to $\neg K \Omega \subseteq \Omega$ for the event $E = \Omega$. We show that the equality $E = \Omega$ is the only case in which the agent's lack of knowledge of an event $E$ is a subset of that same event $E$.

\begin{theorem} \label{t02}
For any event $E \in \mathcal{E}$, if $\neg KE \subseteq E$, then $E = \Omega$.
\begin{proof}
For all events $E \in \mathcal{E}$, Truth implies that the event $\neg KE = \neg E \cup E \setminus KE$ is the union of two disjoint events $\neg E, E \setminus KE \in \mathcal{E}$, $\neg E \cap E \setminus KE = \emptyset$, see Fig. \ref{ss01}. The condition $\neg E \neq \emptyset$ clearly implies $\neg KE \not \subset E$, see Eq. \ref{refine01}. If we find the condition $\neg KE \subseteq E$, considering that we always have $E \setminus KE \subseteq E$, then it necessarily implies $\neg E = \emptyset$, therefore $E = \Omega$.
\end{proof}
\end{theorem}

Our main result derives from the next Theorem \ref{t03}, which shows that the agent's introspection about her lack of knowledge $\neg K \Omega$ is always empty.

\begin{theorem} \label{t03}
$K \neg K \Omega = \emptyset$.
\begin{proof}
Regard the general case $\emptyset \subseteq \neg K \Omega \subseteq \Omega$. Truth implies $K \neg K \Omega \subseteq \neg K \Omega$, but Monotonicity implies $K \neg K \Omega \subseteq K \Omega$. Since the sets are clearly disjoint, $K \Omega \cap \neg K \Omega = \emptyset$, then it implies $K \neg K \Omega = \emptyset$.
\end{proof}
\end{theorem}

We proceed with the analysis. Let the agent's knowledge of the full state-space satisfy Truth, $K \Omega \subseteq \Omega$. Introspection about her knowledge $K \Omega$ provides the event $KK\Omega$, for this event is at most a refinement of her knowledge, $KK \Omega \subseteq K \Omega$. Introspection about her lack of knowledge $\neg K \Omega$ provides $K \neg K \Omega = \emptyset$, which is always empty by Theorem \ref{t03}, so included in $K \Omega$ vacuously. Thinking about every event that the agent either knows or that she does not know provides just the same knowledge which she already has, for we have $K(K \Omega \cup \neg K \Omega) = K \Omega$. It is possible to show that the knowledge of every event $E \in \mathcal{E}$ is included in her knowledge of the full state-space $K \Omega$, see Theorem \ref{t02}\footnote{Formally, the knowledge operator $K$ satisfying Truth and Monotonicity is not necessarily completely additive, i.e. for any events $E,F \in \mathcal{E}$, the weak condition $KE \cup KF \subseteq K(E \cup F)$ applies. We have $KE \cup K \neg E \subseteq K \Omega$.}.

In special, the empty introspection $K \neg K \Omega = \emptyset$ in Theorem \ref{t03} shows that there is not any state in $\Omega$ in which the agent knows the event $\neg K \Omega$. Hence, the agent cannot distinguish between the general case where she lacks knowledge of some event, represented by the non-empty set $\neg K \Omega \neq \emptyset$, vs. the trivial case with $\neg K \Omega = \emptyset$, since introspection about $\neg K \Omega$ provides the same empty outcome in both cases, $K \neg K \Omega = \emptyset$. It means that the agent cannot know if she knows everything or not, i.e. she cannot know if $\neg K \Omega$ is empty or not, therefore she cannot know if $K\Omega$ is equal to $\Omega$ or not.

Moreover, this impossibility is not resolved by learning more about events. For example, regard a very simplified dynamical setting with two sequential stages, $s \in \{0,1 \}$, the operator $K_{s}$ representing the agent's knowledge at the stage $s$. Truth and Monotonicity apply. Assume that at $s = 1$, the agent is capable to process more information about each event $E \in \mathcal{E}$ than at $s = 0$. The condition $K_{0} E \subseteq K_{1} E$ means that given any event $E \in \mathcal{E}$, her knowledge $K_{1} E$ may include more information than her prior knowledge $K_{0} E$.

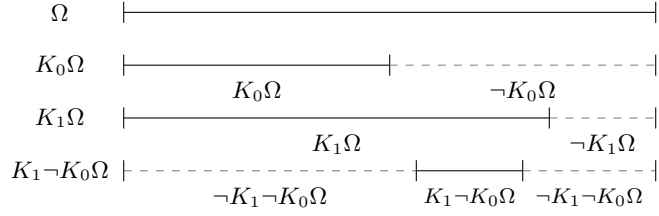
\begin{figure}[htb!]
\begin{center}
\begin{tikzpicture}[scale=0.7,font=\small] 
	\draw (0,10) -- (10,10); 
	\draw (0,10.2) -- (0,9.8); 
	\draw (10,10.2) -- (10,9.8); 
	\node (omega) at (-1.2,10) {$\Omega$}; 
	\draw (0,9) -- (5,9); 
	\draw (0,9.2) -- (0,8.8); 
	\draw (10,9.2) -- (10,8.8); 
	\draw (5,9.2) -- (5,8.8); 
	\draw[gray, dashed] (5,9) -- (10,9); 
	\node (levent) at (-1.2,9) {$K_{0} \Omega$}; 
	\node (event) at (2.5,8.5) {$K_{0} \Omega$}; 
	\node (negevent) at (7.5,8.5) {$\neg K_{0} \Omega$}; 
	\draw (0,8) -- (8,8); 
	\draw[gray, dashed] (8,8) -- (10,8); 
	\node (lkevent) at (-1.2,8) {$K_{1} \Omega$}; 
	\draw (0,8.2) -- (0,7.8); 
	\draw (10,8.2) -- (10,7.8); 
	\draw (8,8.2) -- (8,7.8); 
	\node (kevent) at (4,7.5) {$K_{1} \Omega$}; 
	\node (negkevent) at (9,7.5) {$\neg K_{1} \Omega$}; 
	\draw[gray, dashed] (0,7) -- (5.5,7); 
	\draw (5.5,7) -- (7.5,7); 
	\draw[gray, dashed] (7.5,7) -- (10,7); 
	\node (lknegkevent) at (-1.2,7) {$K_{1} \neg K_{0} \Omega$}; 
	\draw (0,7.2) -- (0,6.8);
	\draw (10,7.2) -- (10,6.8);
	\draw (7.5,7.2) -- (7.5,6.8); 
	\draw (5.5,7.2) -- (5.5,6.8); 
	\node (knegkevent) at (6.5,6.5) { {\footnotesize $K_{1} \neg K_{0} \Omega$} }; 
	\node at (2.75,6.5) {$\neg K_{1} \neg K_{0} \Omega$}; 
	\node at (8.75,6.5) { {\footnotesize $\neg K_{1} \neg K_{0} \Omega$} }; 
\end{tikzpicture}
\end{center}
\caption{An agent's knowledge of the full state-space at stages $s \in \{0,1 \}$, $K_{s} \Omega$.} \label{ss02}
\end{figure}

Begin at $s = 0$. Let the agent learn that some event $E$ obtains, so her knowledge of $E$ changes from $K_{0} E = \emptyset$ to $K_{1} E \neq \emptyset$, condition $K_{0} E \subseteq K_{1} E$ applies. By learning about the previously unknown event $E$, the agent knows now (at $s = 1$) that she did not know everything before knowing $E$ (at $s = 0$), otherwise $E$ would not be unknown at $s = 0$. Her epistemic operator after learning about $E$ becomes $K_{1}$, for it provides the introspection about her prior lack of knowledge equal to $K_{1} \neg K_{0} \Omega \neq \emptyset$, which is non-empty\footnote{In detail, at $s = 0$, Truth implies $K_{0} E \subseteq E$, and Monotonicity implies $K_{0} E \subseteq K_{0} \Omega$, therefore we have $K_{0} E \subseteq E \cap K_{0} \Omega$. The lack of knowledge of $E$ such that $E \cap K_{0} \Omega = \emptyset$ provides the events $K_{0} E = \emptyset = E \cap K_{0} \Omega$, $\neg K_{0}E = \Omega = \neg E \cup \neg K_{0} \Omega$. It necessarily implies $E \subseteq \neg K_{0} \Omega$. Now, let the agent learn at $s = 1$ that $E$ obtains, $K_{1} E \neq \emptyset$. Monotonicity implies $K_{1} E \subseteq K_{1} \neg K_{0} \Omega$, therefore $K_{1} \neg K_{0} \Omega \neq \emptyset$.}. This case is represented in Fig. \ref{ss02} where we have a state-space $\Omega$, the agent's knowledge of the full state-space $K_{s} \Omega$ at stages $s \in \{0,1 \}$, and the event $K_{1} \neg K_{0} \Omega$. Truth and Monotonicity imply the condition $K_{1} \neg K_{0} \Omega \subseteq K_{1} \Omega \cap \neg K_{0} \Omega $.


The non-empty introspection $K_{1} \neg K_{0} \Omega \neq \emptyset$ indicates that the agent knows that she did not know everything at stage $s = 0$. Nonetheless, she cannot know if she just learned everything that was left to know, or if she still lacks knowledge of some event at the current stage $s = 1$. Condition $K_{1} \neg K_{0} \Omega \subseteq K_{1} \Omega$ indicates that the agent's knowledge of her prior lack of knowledge is included in her current knowledge of the full state-space, $K_{1} \Omega$, for the non-empty complement $\neg K_{1} \Omega \neq \emptyset$ still may hold, see example in Fig. \ref{ss02}.

Our analysis is similar to the one in the static case. Introspection about her knowledge $K_{1} \Omega$ provides the event $K_{1} K_{1} \Omega$, which is included in $K_{1} \Omega$ following Truth. Introspection about her lack of knowledge $\neg K_{1} \Omega$ provides $K_{1} \neg K_{1} \Omega = \emptyset$, which is always empty by Theorem \ref{t03}. Thinking about everything that she either knows or that she does not know provides the same knowledge which she has after learning about the last unknown event, equal to $K_{1}(K_{1}\Omega \cup \neg K_{1} \Omega) = K_{1} \Omega$. The empty introspection $K_{1} \neg K_{1} \Omega = \emptyset$ derived in Theorem \ref{t03} indicates that the agent cannot know if her lack of knowledge $\neg K_{1} \Omega$ is empty or not, therefore she cannot know if her knowledge $K_{1} \Omega$ is equal to $\Omega$ or not. It shows that the agent still cannot know if she knows everything or not.

\section{Discussion} \label{Discussion}

We show that a rational agent with true and refinable knowledge of events cannot know if she knows everything or not. We assume that the agent's knowledge is bounded by her epistemic operator $K$, which represents her capacity to process information about events. This epistemic condition means that some states of the world may be inaccessible by the agent, i.e. in modal logic, it means that the accessibility relation is not reflexive with respect to each state. Consequently, the baseline property of Necessitation is not satisfied. Notwithstanding, literature shows that model systems characterised by topological semantic structures and satisfying conditions corresponding to Truth and Monotonicity (as our model) are sound and complete.

Remark that Truth and Monotonicity imply the condition $KE \subseteq E \cap K \Omega$, under which the set $K \Omega$ engages in a similar position as the so-called "awareness set" found in the literature. Set-algebraic models that assume Necessitation often require an awareness operator, in addition to the knowledge operator, in order to avoid inconsistencies in modelling the lack of knowledge and the unawareness of agents. We provide conditions which may enable knowledge to be a homomorphic representation of awareness, therefore simplifying current epistemic models while preserving consistency.



\end{document}